\documentclass[aps,prb,twocolumn,showpacs,superscriptaddress,groupedaddress]{revtex4}

\usepackage{graphicx}
\usepackage{dcolumn}
\usepackage{bm}
\usepackage{amsmath}
\usepackage{epstopdf}

\usepackage{bm}
\begin{document}
\title{Spin waves in one-dimensional bi-component magnonic quasicrystals}

\author{J.~Rych{\l}y}
\email{rychly@amu.edu.pl}
\author{J.~W.~K{\l}os} \email{klos@amu.edu.pl}
\author{M.~Mruczkiewicz}
\author{M.~Krawczyk} \email{krawczyk@amu.edu.pl}
\affiliation{Faculty of Physics, Adam Mickiewicz University, Umultowska 85, Pozna\'{n}, 61-614, Poland}
\date{\today}

\begin{abstract}
We studied finite Fibonacci sequence of Co and Py stripes aligned side-by-side and in direct contact with each other. Calculations based on continuous model including exchange and dipole interactions were performed for structures feasible for fabrication and characterization of main properties of magnonic quasicrystals. We have shown the fractal structure of the magnonic spectrum with a number of magnonic gaps of different widths. Also localization of spin waves in quasicrystals and existence of surface spin waves in finite quaiscrystal structure is demonstrated. 
\end{abstract}

\maketitle
\section{Introduction}
Quasicrystals\cite{Levine,Janot} are aperiodic structures with long range order which can be constructed in deterministic way by self-similar replication or by partial projection of appropriate periodic structure from higher dimensions.\cite{Vardeny13} This symmetry can be revealed in diffraction pattern showing the set of discrete diffraction spots. The diffraction pattern can be described by the structure factor which, for quasiperiodic lattice, consists of dense set of delta peaks. The location and the amplitude of peaks of the structure factor is, on the other hand, related to the position and width of frequency/energy gaps.\cite{Janot} This feature of quasicrystals allows for advanced tailoring of the band structure which exceeds possibilities offered by homogeneous material or artificial crystals.\cite{Macia06,Janssen14}

One-dimensional (1D) quasicrystals in the form of Fibonacci sequence have already been studied for photonic,\cite{Vardeny13,Kohmoto87a} electronic\cite{Merlin85,Laruelle88} and phononic\cite{Steurer07} systems. However, research of spin waves (SWs) in 1D magnonic quasicrystals (MQs) is so far mainly limited to theoretical investigations of multilayered structures.\cite{Albuquerque03} Attention was focused on lattice models for exchange SWs\cite{Costa13,Liu93} and in continuous model for magnetostatic SWs.\cite{Anselmo99}
Experimentally, SW excitations in quasicrystals were investigated only recently regarding their localization properties in two-dimensional structures made of thin veins of Py\cite{Bhat13,Bhat14} and in a larger scale for enhancement of nonlinear effects in thick yttrium-iron-garnet film with Fibonacci sequence of etched groves. \cite{Grishin13} These studies have shown interesting properties of MQs potentially useful for applications and fundamental studies.
However, there is an evident gap in studies of SW dynamics in MQs, especially in studies of structures feasible for experimental realization. 
This area can be explored theoretically by considering finite structure of planar geometry and including both, exchange and dipole interactions. Therefore, we focus our study on planar bi-component MQs modulated in nanoscale, i.e., on structures preserving all features characteristic for wave dynamics in quasiperiodic structures which are possible for fabrication and characterization.

We considered a thin plate of 1D MQ in the form of finite Fibonacci sequence consisting of Co and permalloy (Py: Ni$_{80}$Fe$_{20}$) stripes. Ferromagnetic stripes were in direct contact which ensures exchange coupling between neighboring stripes, additional to long range dipole interactions. We investigated system in saturation state with static magnetization aligned along infinite stripes. 
The strong dynamical coupling makes system dispersive for propagating SWs with interesting features like a fractal SW spectra and localized properties of SWs. Moreover,
the finite Fibonacci sequence allowed us to study surface SWs localized at edges of the MQ. To find frequencies of SWs and their spatial profiles we solve linearized Landau-Lifshitz equation with finite element method (FEM) in frequency domain.\cite{Mruczkiewicz13b,Mruczkiewicz13}

First we present the structures under investigation and describe shortly the used computational method (Sec.~\ref{sec:model}). In the Sec.~\ref{sec:res} we present results of our calculations and perform discussion of obtained results. Finally, in the last section, we summarize the paper.  

\section{Model and the structure}\label{sec:model} 
\begin{figure}[!ht]
	\includegraphics[width=7cm]{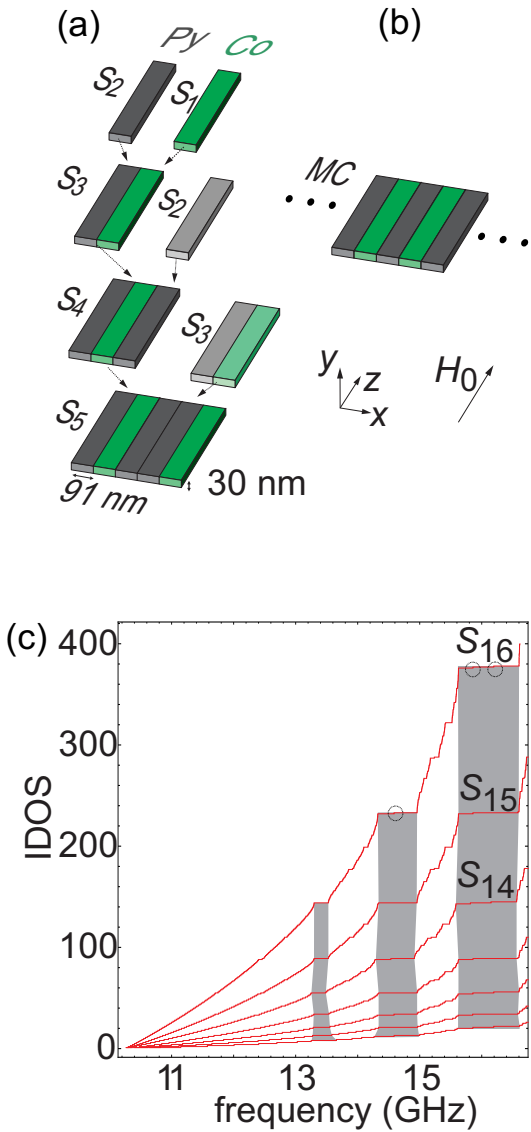}
	\caption{(Color online)(a) First few sequences of planar Fibonacci MQ (Fibonacci sequences from $S_1$ to $S_5$ are shown) composed of Py and Co stripes of finite width and thickness. (b) 1D bi-component MC with Py and Co stripes of the same width. Both structures are saturated by external magnetic field $H_0$. (c) Integrated density of states (IDOS) as a function of frequency for successive Fibonacci sequences: from $S_{10}$ (bottom curve) to $S_{16}$ (top curve). In the IDOS of $S_{16}$ MQ three surface SWs existing in widest magnonic gaps are marked with circles.} \label{Fig:struc}
\end{figure}

We have investigated SW spectra for 1D MQs obtained according to Fibonacci inflation rule and made from long Co and Py stripes of 91 nm width and 30 nm thickness. Saturation magnetization and exchange constant for Co and Py are: $M_{\text{Py}}=0.86\times 10^{6}$ A/m, $A_{\text{Py}}=1.3\times 10^{-11}$ J/m, $M_{\text{Co}} = 1.445\times 10^{6}$ A/m, $A_{\text{Co}}=3.0\times 10^{-11}$ J/m.\cite{Wang09,Sokolovskyy11} For each material gyromagnetic ratio $\gamma=176$ rad GHz/T is assumed the same. In accordance with the Fibonacci inflation rule, Co and Py stripes are arranged in Fibonacci sequence using the following recursion: $S_{n}=[S_{n-1}S_{n-2}]$ \cite{Kohmoto87b}, where $S_1$ and $S_2$ are initial structures consisting of single Co and Py stripe, respectively. $[S_{n-1}S_{n-2}]$ means concatenation of the two subsequences $S_{n-1}$ and $S_{n-2}$ of the stripes. We obtained in first subsequences: Co, Py, PyCo, PyCoPy, PyCoPyPyCo [Fig. \ref{Fig:struc}(a)]. In each next step the structure total width is growing in $x$ direction, e.g., for the quasicrystal made from 55 stripes the width of the whole structure is 5 $\mu$m. We have analyzed structures $S_{10}$, $S_{11}$, $\cdots$ $S_{16}$, made of 55, 89, 144, 233, 377, 610 and 987 number of stripes. With $n\rightarrow\infty$ the filling fraction of Py approach to the golden ratio $\sigma=(\sqrt{5}-1)/2$ and the sequence of stripes $S_{\infty}$ becomes rigorously quasiperiodic. We use the coordinate system as defined in Fig.~\ref{Fig:struc}. Along stripes structure is infinite and saturated by external magnetic field $\mu_0H_0 = 0.1$ T. To emphasize interesting properties of SWs in quasicrystals we will make reference to magnonic spectra calculated for MC composed of the same stripes of Py and Co as Fibonacci sequence [Fig. \ref{Fig:struc}(b)].

To calculate SW spectra we solve the Landau-Lifshitz equation (LLE):
\begin{eqnarray} 
	\frac{\partial {\bf M}({\bf r},t)}{\partial t}&=&\mu_0\gamma [ {\bf M}({\bf r},t)\times{\bf H}_{\rm eff}({\bf r},t) \nonumber \\
	&+& \frac{\alpha}{M_{\mathrm{S}}} {\bf M}({\bf r},t) \times \left( {\bf M}({\bf r},t) \times {\bf H}_{\rm eff}({\bf r},t)\right)],\label{eq:LLE}
\end{eqnarray}
where $t$ is time and ${\bf r}$ is position vector. The last term describes damping of SWs with $\alpha$ being dimensionless damping coefficient. ${\bf H}_{\rm eff}$ is effective magnetic field, which is assumed to be the sum of three terms: ${\bf H}_{\rm eff}={\bf H}_{\rm 0}+{\bf H}_{\rm ex}+{\bf H}_{\rm dm}$. ${\bf H}_{\rm ex}$ is exchange field and ${\bf H}_{\rm dm}$ is dynamic demagnetizing field with components along $x$ and $y$ directions (due to assumed geometry the static demagnetizing field is 0). The ${\bf H}_{\rm ex}$ and ${\bf H}_{\rm dm}$ fields are defined in Ref.~[\onlinecite{Mruczkiewicz13b}]. We neglect magnetic anisotropy term in ${\bf H}_{\rm eff}$, because its influence on presented results in MQs and MCs composed of Py and Co nanostripes is small.\cite{note_10} 

From Eq. (\ref{eq:LLE}) we find dynamical components of magnetization, ${\bf m}({\bf r},t)$ [${\bf M}({\bf r},t)= M_{z}({\bf r}){\bf e}_{z}+{\bf m}({\bf r},t)$]. We use linear approximation, i.e., we neglect the higher order terms arising in Eq. (\ref{eq:LLE}) with respect to ${\bf m}$. This is justified when ${M}_{z}$ is assumed to be constant in time, namely when $|{\bf m}({\bf r},t)|\ll M_{z}{\bf e}_{z}$, and therefore ${M_z}\approx M_{\mathrm{S}}$ ($M_{\mathrm{S}}$ is saturation magnetization). We seek solutions of LLE (\ref{eq:LLE}) in the form of monochromatic SWs, having harmonic dynamics in time: $e^{i\omega t}$, where $\omega$ is angular frequency of SW. Eq. (\ref{eq:LLE}) is complemented with Maxwell equations to determine demagnetizing fields. With these equations, we define the eigenvalue problem, which is solved by using FEM with COMSOL 4.3a software to obtain dispersion relation and profiles of SWs. For more details concerning this computation we refer to Ref.~[\onlinecite{Mruczkiewicz13}]. From solution of Eq.~(\ref{eq:LLE}) we found complex values of $\omega$ with its real part corresponding to SW frequency and imaginary part proportional to the inverse of time of life of SW.\cite{Gurevitch} We have checked numerically, that the influence of damping on SW frequencies (by assuming $\alpha$ coefficients 0.01 and 0.1 for Py and Co, respectively, in one calculations and 0 in other) is smaller than 0.5.\cite{note_11} Thus, in further calculations we neglect damping.



\section{Results and discussion}\label{sec:res}
To visualize the SW spectra we use integrated density of states IDOS, defined as:
\begin{equation}
	{\rm IDOS}(f_i)=\sum_{j=0}^{i}{\rm DOS}(f_j),\label{eq:IDOS}
\end{equation}
where DOS is density of SW modes and $f_i$ is a frequency of the $i$-th SW mode, which are ordered according with increased frequency.\cite{note_33} IDOS as a function of frequency calculated for successive Fibonacci sequences from $S_{10}$ to $S_{16}$ are shown in Fig. \ref{Fig:struc}(c).

Due to the finite width of structures used in calculations, we always observe discrete set of frequencies. The finer steps and faster increase of IDOS are observed for wider structures (i.e. composed of larger number of nanostripes). By finding plateaus in IDOS we are able to identify magonic gaps. The width of these plateaus converges for larger Fibonacci structures and can approximate width of magnonic gaps in infinite 
MQs [see the gray areas marked in Fig.~\ref{Fig:struc}(c)]. Within some plateaus of IDOS, we can also find surface modes of MQ. In IDOS shown in Fig. \ref{Fig:struc}(c) these surface SWs are indicated by separated steps (marked with circles for $S_{16}$) inside the magnonic gap, discussed in further part of the paper.

\begin{figure}[!ht]
	\includegraphics[width=7cm]{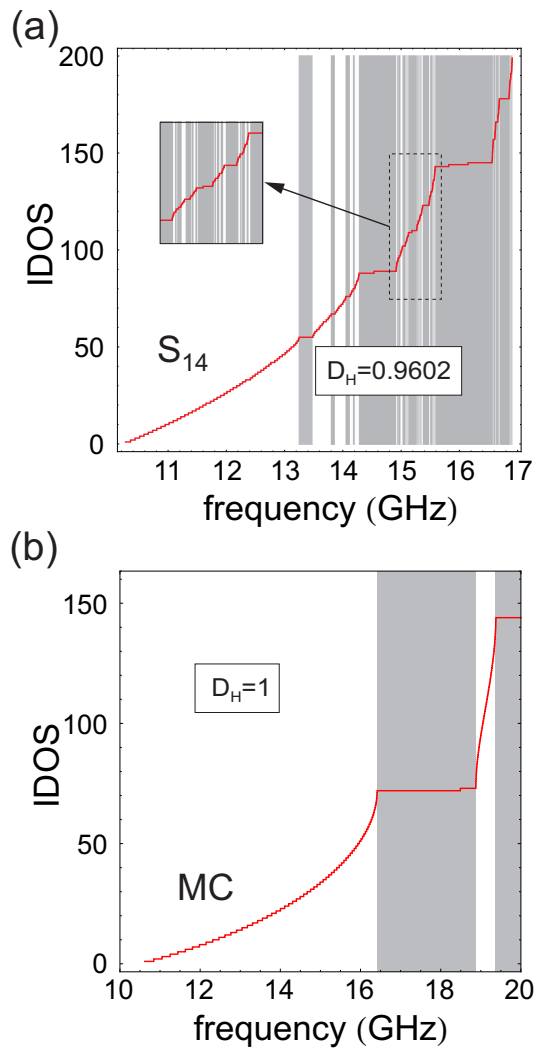}
	\caption{(Color online) (a) IDOS for Fibonacci sequence $S_{14}$. Gray areas mark the most pronounced magnonic gaps. Inset presents magnified region of IDOS plot where the complex bandgap structure is visible. The Hausdorff dimension $D_{\rm H}$ is 0.9602. (b) IDOS for MC consisting of 144 nanostripes, $D_{\rm H} = 1$.} \label{Fig:DH}
\end{figure}

With the increase (decrease) of size of the structure, the spectrum of IDOS reveals more (less) details of spectrum characteristic for quasicrystals (see Fig.~\ref{Fig:struc}(c)). For larger structures, we are able to identify the fine structure of magnonic gaps, whereas gaps found for shorter sequences still exist.\cite{note_44} This provides mechanism to explore fractal nature of SW spectrum in MQs. In Fig.~\ref{Fig:DH}(a) we present in details the spectrum for structure consisting of 377 nanostripes. The inset in this figure shows magnified region of spectrum with subtle, multilevel structure of magnonic gaps, resembling property of self-similarity. 

In order to validate the fractal nature of SW spectrum we have calculated its Hausdorff dimension.\cite{Zaginaylov02} We divided the whole investigated frequency range into intervals of equal lengths $\Delta f$ and then we counted the number $N(\Delta f)$ of these intervals which are included in or partially overlap with magnonic bands. The number $N(\Delta f)$ increases with decrease of length $\Delta f$ and dependence of $\log\left[N\left(\Delta f\right)\right]$ on $\log\left(\frac{f_0}{\Delta f}\right)$ should be linear ($f_0$ is a frequency of the first mode in spectra, its choice is arbitrary and does not influence results). The Hausdorff dimension of the spectrum is defined as the derivative:
\begin{equation}
	D_{\rm H}=\frac{{\rm d}\, \log\left[N\left(\Delta f\right)\right]}{{\rm d}\,\log\left(\frac{f_0}{\Delta f}\right)}.\label{eq:DH}
\end{equation}
Numerically we have calculated $D_{\rm H}$ as a coefficient of regression for dependence of $\log\left[N\left(\Delta f\right)\right]$ on $\log\left(\frac{f_0}{\Delta f}\right)$. We have obtained value $D_{\rm H}$ = 0.9603 with the standard deviation 0.0011 and the coefficient of regression $R^2$ = 0.99991, which points close to linear dependence. This non integer value of $D_{\rm H}$ proves fractal property of SW spectra in MQ. Its value is close to the $D_{\rm H}$ obtained for plasmonic Fibonacci structures.\cite{Zaginaylov02} In the considered range of frequencies (which is accessible for experimental investigations) we found the $D_{\rm H}$ practically independent on the size of intervals $\Delta f$. This indicates that the spectrum is not multifractal one and can be characterized by single $D_{\rm H}$ dimension.

IDOS for MCs has a regular dependence on $f$. 
In Fig.~\ref{Fig:DH}(b) we have plotted the IDOS$(f)$ for MC structure consisting of 144 nanostripes.
Here, $D_{\rm H}$ is equal 1. The first magnonic band gap starts at 16.5 GHz, however in MQ gaps are present already at lower frequencies. Existence of low frequency gaps in MQs can be useful for application, for instance in magnon transistors.\cite{Chumak2014}

The other interesting issue of excitations in quasicrystals is a possibility for their spatial localization in MQ interior.\cite{Kohmoto87a} In crystals without any defects, all modes are extended, but in random systems localization occur.\cite{Abrahams1979,Wolf1985} MQs are neither periodic nor random systems. Therefore, we can observe both extended and localized SW modes.

To discuss localization quantitatively, we need measure of the strength of localization. For this purpose we will use localization factor $\lambda$.\cite{Xie00} The parameter $\lambda_{i}$ for $i^{\rm th}$ mode is defined as:
\begin{equation}
	\lambda_{i}=\frac{1}{L}\sum_{j=0}^{j=L}\log\left|m_{x,i}\left(x_{j}\right)\right|,\label{eq:lambda}
\end{equation}
where the in-plane dynamical component of magnetization $m_{x,i}\left(x_{j}\right)$ is taken at the point $x_{j}$ and is normalized according with the norm:
\begin{equation}
	\frac{1}{L}\sum_{j=0}^{j=L}\left|m_{x,i}\left(x_{j}\right)\right|=1.\label{eq:lambdaN}
\end{equation}
Summation runs over all $L$ equidistant points $x_{\rm j}$ covering all the space between surfaces of structure along $x$-axis. The $\lambda=0$ corresponds to uniform excitation. For all other modes $\lambda$ takes negative values. Modes with stronger localization are characterized by large absolute value of $\lambda$. 

In the considered systems, the separation of frequencies of SW modes quantized across the thickness exceeds the frequency range considered in this paper. Thus, for further analysis of localization factors we take SW amplitude from the middle plane of the structure.

\begin{figure}[!ht]
	\includegraphics[width=7cm]{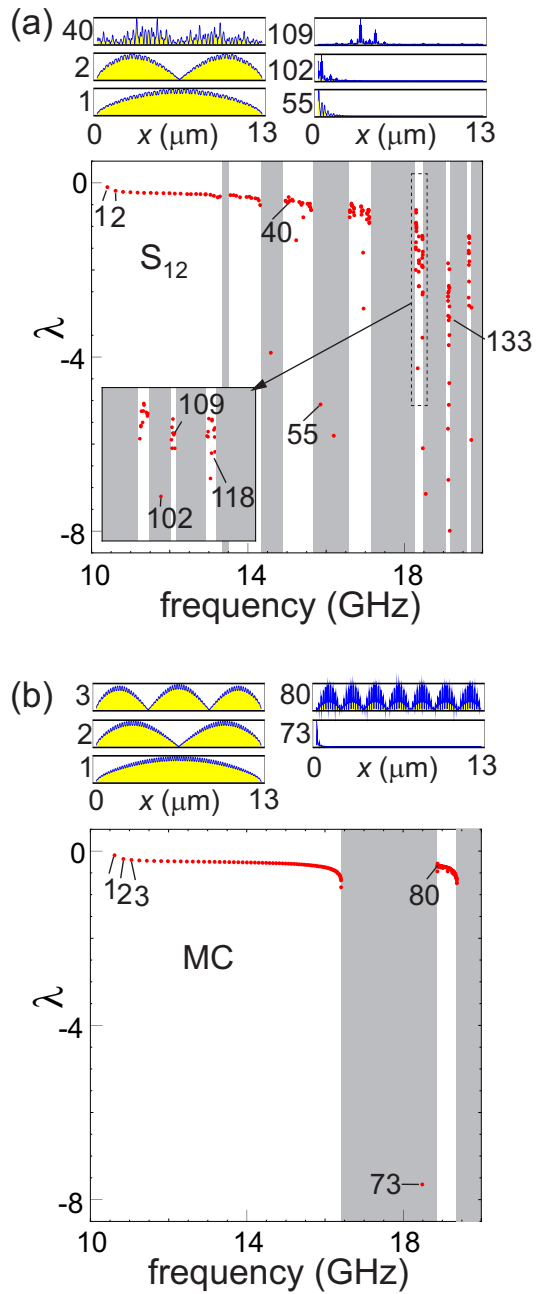}
	\caption{(Color online) Localization factors $\lambda$ of SW modes in (a) $S_{12}$ sequence of the Fibonacci MQ and (b) of MC composed of the same number of nanostripes in dependence on frequency. Numbers labeling points in the $(\lambda, f)$ plot denote successive numbers of modes ordered with increased frequency. Profiles of the modulus of $x$-component of magnetization along the structure are plotted in insets above the main plot. Region of selected band in (a) is magnified in separate inset. Grey areas mark magnonic gaps of the widest width.} \label{Fig:lambda}
\end{figure}	

\begin{figure}[!ht]
\includegraphics[width=7.5cm]{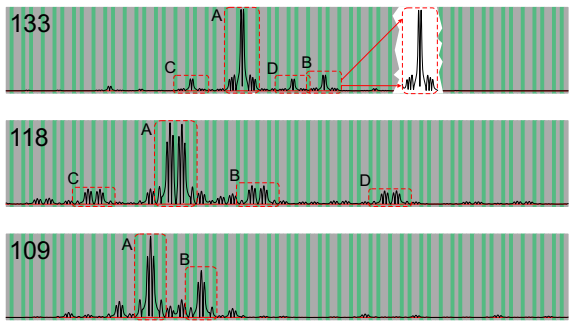}
\caption{(Color online) The profiles of the modulus of the $x$ component of magnetization in $S_{12}$ Fibonacci MQ for the bulk modes no. 109, 118 and 133, labeled in Fig.~\ref{Fig:lambda}(a). The gray and green bars in the background mark the regions of Py and Co, respectively. The letters A,B,C and D point the areas (marked with dashed lines) of SWs localization with similar surroundings.} \label{Fig:profiles}
\end{figure}

Localization factors of SWs in MC and the Fibonacci quasicrystal ($S_{12}$) are shown in Fig. \ref{Fig:lambda}. In general, the localization factor $\lambda$ increases in magnitude with growing frequency. For extended modes (e.g., in MCs or low frequency part of the MQs spectra) with increasing frequency, the number of nodal points in the profiles of SWs increases (see the profiles of SWs in Figs.~\ref{Fig:lambda}). Because Py has lower ferromagnetic resonance frequency than Co, the modes distribute their amplitude more in Py stripes with nodal points placed mainly in Co stripes. Such inhomogeneous distribution increases the magnitude of $\lambda$. The rate of this change is relatively small for MC (see Fig.~\ref{Fig:lambda}(b)) where the localization factor for SWs up to 20 GHz does not fall below -1 (apart from the mode no. 73 being a surface SW discussed later). The changes of localization factors in the MQ ($S_{12}$ shown in Fig.~\ref{Fig:lambda}(a)) have similar tendency up to $\approx 15$ GHz.

For higher frequencies modes oscillate more rapidly in space and feel the structure stronger. Fibonacci sequence is much more complex than periodic one, thus the high frequency excitations are essentially different for MQs than for MCs. There are three reasons for this: (i) in MQ are present Py stripes of both single and double width, 91 and 182 nm, respectively (see, Fig.~\ref{Fig:struc}(a)); (ii) the Py and Co stripes are distributed quasiperiodically within the MQ and (iii) the neighborhood of each stripe can be different in MQ, while it is always the same in MC. In the considered, higher frequency range, the modes are concentrated in selected (single or double) Py stripes in MQs. First group of modes above extended mode range, consist of modes with amplitude concentrated in some Py stripes of the double width with one nodal point in each stripe. Then there is a group of modes with two nodal points in the double Py stripe, i.e., a second harmonic of the double stripe (e.g. mode no. 109 and no. 118 in Fig. 4) and then group of modes concentrated in a few single Py stripes with one nodal point within the stripe, e.g. mode no. 133.

Each of the discussed above localized bulk modes is located at given sets of positions (single or double Py stripes), mostly because of the similarities between surroundings of positions of SW amplitude localizations. For selected frequency, at most few Py stripes with very similar soundings are able to hold the same excitation. In Fig.~\ref{Fig:profiles} we analyze in details selected profiles of those bulk modes (the modes of numbers: 109, 118, 133), localized within the $S_{12}$ MQ, being the modes of higher frequencies (18.37 GHz, 18.47 GHz and 19.13 GHz, respectively). The highest amplitude for the mode no. 109 is observed in the double width Py stripes. We mark the two locations A and B with the highest amplitude of SWs. By zooming in less intense excitation at location B we noticed that it has the same spatial profile as the most intense excitation from location A. This similarity can be explained by inspecting the neighborhood of these two locations. For the location A the following sequence of single ($s$) and double ($d$) width Py stripes can be found: $|$${\bf s}dsddsdsddsdds$A$sddsdsddsddsdsdsddsd{\bf d}$… The symbol ‘$|$’ denotes the surface of the structure. By swapping the left and the right surroundings of B the following sequence of stripes could be found: $ds{\bf d}dsddsdsddsdds$B$sddsdsddsddsdsdsddsd{\bf s}$. It is easy then to notice that the sequences of stripes in the neighborhood of location A and B differ at very far positions (i.e. at positions 14th and 19th counting from the position of A and B), which are marked with bold letters in the sequences. The similar discussion can be done for the modes no. 118 and no. 133. It also could be noted that the mode no. 118 is concentrated in the two double Py stripes separated by the single Co stripe, whereas the mode no. 133 is focused within one single Py stripe.

It was shown that for photonic and electronic structures the localization of the modes in Fibonacci structures is supposed to be weaker than exponential and is governed by power law dependence.\cite{Negro,Capaz} This property makes them substantially different from defect modes in disordered structures resulting from Anderson localization or surface modes in periodic systems induced by termination of the structure. We expect that this 'chaotic' modes \cite{Kohmoto87b} will have self-similar properties\cite{Capaz} also for magnonic systems.

We discuss now the modes with amplitude localized at the surface of the structure, i.e., the surface states, found in MCs and MQs spectra. Occurrence of surface modes is determined mostly by the element placed on the surface of the crystal (in our case it is the single stripe of Py) and its frequency overlapping with the gap.\cite{Steslicka}  Only up to two surface modes structured on the basis of the excitation with fixed quantization in the surface stripe (i.e., with fixed number of nodes in the Py stripe) can occur. SW spectrum of the finite Fibonacci structure contains many surface modes which exist in frequency gaps, appearing in large number in comparison to periodic structure. Additionally, it is important to notice that the amplitude of surface waves penetrate to the inside of the structure, which also has impact on properties of excitation. Because of that, surface modes appearing in MC and MQ could be different, regardless the same surface element and amplitude distribution in this element. In MC we have found only one surface mode up to 20 GHz [mode no. 73 in Fig.~\ref{Fig:lambda}(b)] but in MQ, many surface modes can be identified [e.g., mode no. 55 shown in Fig.~\ref{Fig:lambda}(a)]. They have $\lambda<-4$, which exceeds the factor of bulk modes. For some modes with frequency near to the end of bunch of bulk modes (e.g. mode no. 102) distinguishing between bulk and surface character can be ambiguous. We can point out that, the strength of localization of surface modes increases if its frequency is located in the center of the gap, when this gap is wider, and when its frequency is high.

Finally we discuss long wavelength part of the spectra. In the low frequency limit, both periodic and Fibonacci structures exhibit similar properties: much the same profiles [see insets in Fig. \ref{Fig:lambda}(a) and (b) for modes 1 and 2], frequencies and localization factors of the SWs [compare Fig.~\ref{Fig:lambda}(a) and \ref{Fig:lambda}(b)]. In this limit both systems can be treated as metamaterials characterized by effective magnetic properties.\cite{Mruczkiewicz12} It is worth noting, that in this limit, the ${\rm IDOS}(f)$ spectra of MQs and MCs have features characteristic for the Damon-Eshbach wave in homogeneous film.\cite{Wang09,Sokolovskyy11}

\begin{figure}[!ht]
	\includegraphics[width=7cm]{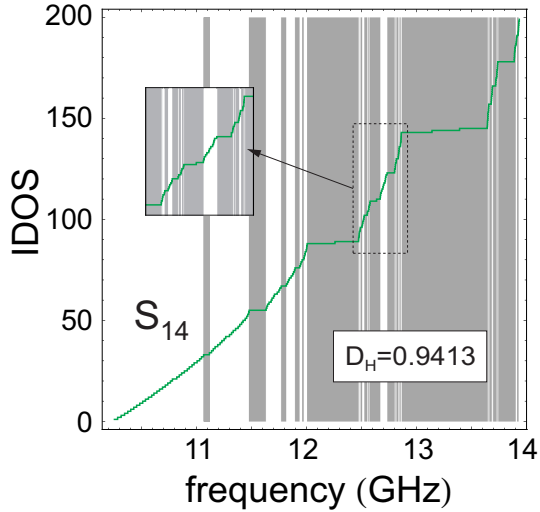}
	\caption{(Color online) IDOS of Fibonacci sequence $S_{14}$ composed of wide Py and Co stripes--250 nm width and 30 nm thickness. Gray areas mark the most pronounced magnonic gaps. Inset presents the magnified region of IDOS plot where the complex bandgap structure is visible. Hausdorff dimension of the spectrum is $D_{\rm H}= 0.9413$.} \label{Fig:250}
\end{figure}

We have proposed the MQ composed of Py and Co stripes suitable for experimental study of fractal properties in magnonics. Indeed, magnonic band structure have already been investigated in MCs composed of Py or Co and Py stripes with periodicity $\propto$ 500 nm and interesting physical properties, like magnonic band gaps and re-programmability have been demonstrated.\cite{Wang09,Topp2010,Ding11,Tacchi12} The Brillouin light scattering (BLS) to measure dispersion relations of SWs, micro-BLS spectroscopy\cite{Madami12} and time resolved magneto-optical Kerr effect (TR-MOKE) microscopy\cite{Au2011,Barman14} with spatial resolution down to 250 nm, to visualize profiles of SW excitations, have been used. These techniques can be directly implemented to study SWs in MQs and to confirm predicted properties. The area of SW localization in MQs, under investigation, spreads over few Py stripes [e.g., modes 55, 102 and 109 in Fig.~\ref{Fig:lambda}(a)], which is above limits of spatial resolution in BLS and TR-MOKE microscopes. Such direct studies of SW localization are extremely difficult for layered systems but can be performed for considered structures of in-plane periodicity. Moreover, properties of SWs in MQs presented above preserve also in larger structures, i.e., with stripe width extended up to few hundreds of nm. To confirm this we present in Fig.~\ref{Fig:250} IDOS as a function of frequency calculated for MQ composed of 250 nm width Py and Co stripes i.e., the size which exactly matches stripes of MCs studied in Ref.~[\onlinecite{Wang09}]. The spectra are at lower frequencies than MQs with stripes of 90 nm width [Fig.~\ref{Fig:DH}(a)], but has the same fractal structure confirmed by slightly smaller Hausdorff dimension $D_{\text{H}}=0.9413$.

\section{Summary}
We have investigated theoretically periodic and quasiperiodic planar magnonic systems suitable for experimental realization. Systems consisting of thin Py and Co stripes arranged in periodic and Fibonacci structures where considered. We have shown that the spectrum of IDOS for MQs systems exhibits complex, multilevel structure of frequency gaps with finer details revealed for long Fibonacci sequences. Calculated magnonic spectra form fractal set with self-similarity property characterized by the Hausdorff dimension 0.9603. 
Computed localization factors allow us to discuss quantitatively the strength of localization for SWs in MQs. We have shown the localization of bulk modes in quasiperiodic systems and that it is enhanced for higher frequencies, where spectrum of Fibonacci structures becomes substantially complex. The presence of multiple magnonic gaps supports also the existence of surface SWs which are observed numerously in MQs. In the long wavelength limit both MQs and MCs preserve similar effective properties. Obtained results show that quasicrystal structures can be investigated in magnonics with standard experimental techniques and their fractal properties can be explored. 

\section*{Acknowledgements}
	\textbf{Acknowledgements}. The authors would like to thank Dr Andriy E. Serebryannikov for fruitful discussions. The presented research has received funding from Polish National Science Centre project DEC-2-12/07/E/ST3/00538, project SYMPHONY (program TEAM/2011-8/4) of the Foundation for Polish Science and from the EU’s Horizon2020 research and innovation programme under the Marie Sk{\l}odowska-Curie GA No644348.


\begin{thebibliography}{xx}
\bibitem{note_10} According to experimental studies in MCs composed of polycrystalline Co and Py nanostructures deposited with electron beam lithography magnetocrystalline and magnetic anisotropy from the interface with the substrate is small and have negligible effect on SW dynamics.\cite{Wang09,Tacchi12} Thus, in our calculations devoted to similar structures we have neglected magnetic anisotropies.
\bibitem{note_11} The imaginary part of $\omega$ is closer to the respective value of homogeneous Py film than of Co film for SWs investigated in this paper, because the main contribution to these excitations comes from Py. Similar as in planar MCs,\cite{Wang09} this damping shall not influence significantly experiments proposed at the end of this paper.
\bibitem{note_33} IDOS defined according with Eq. (\ref{eq:IDOS}) is in fact the sum of the number of SW frequencies up to the frequency $f_i$.
\bibitem{note_44} In Fig.~\ref{Fig:struc}(c), we marked magnonic gaps (between frequencies of the last and the first bulk mode, in the band below and above given gap, respectively). Width of these frequency ranges can by slightly overestimated due to quantization of frequencies.
\end{thebibliography}
\end{document}